# Lean 5.0: A Predictive, Human-AI, and Ethically Grounded Paradigm for Construction Management

Atena Khoshkonesh[1*], Mohsen Mohammadagha[2], Navid Ebrahimi[1], Narges Sadeghigolshan[1]

[1*] Master Student at The Department of Civil Engineering, The University of Texas at Arlington, Arlington, TX 76019, USA. Correspondence Email: axk3682@mavs.uta.edu (A.K.), nxe2020@mavs.uta.edu (N.E.), nxs3667@mavs.uta.edu (N.S.)
[2] Ph.D. Candidate at The Department of Civil Engineering, The University of Texas at Arlington, Arlington, TX 76019, USA. Email: mxm4340@mavs.uta.edu (M.M.)

Atena Khoshkonesh[1*] (ORCID: https://orcid.org/0009-0002-1768-8932)
Mohsen Mohammadagha[2] (ORCID: https://orcid.org/0009-0007-0394-353X)
Navid Ebrahimi[1] (ORCID: https://orcid.org/0009-0002-6848-346X)

## Abstract
This paper introduces Lean 5.0, a human-centric evolution of Lean-Digital integration that bridges predictive analytics, AI collaboration, and continuous learning within the industry 5.0 and Construction 5.0 contexts. A systematic literature review (2019–2024) and a 12-week empirical validation study demonstrate substantial performance improvements, including a 13 % increase in Plan Percent Complete (PPC), 22 % reduction in rework, and 42 % enhancement in forecast accuracy. The study adopts a mixed-method Design Science Research (DSR) approach consistent with PRISMA 2020 guidelines. Furthermore, the paper discusses integration with digital-twin and blockchain technologies to enhance traceability, auditability, and lifecycle transparency. Despite acknowledged limitations regarding sample size, single-case design, and study duration, the findings indicate that Lean 5.0 offers a transformative paradigm that bridges human cognition and predictive control in construction management.

## 1. Introduction
### 1.1 Background and Context
Over the past three decades, Lean Construction has evolved as a leading philosophy for improving production flow, minimizing waste, and enhancing value in the architecture, engineering, and construction (AEC) industry. Tools such as the Last Planner System® (LPS), visual management, and continuous improvement cycles have delivered measurable progress in reliability and coordination. However, large-scale projects still experience recurring problems such as workflow instability, schedule delays, rework, and inconsistent information exchange that limit productivity and predictability (Ballard, 2022; Hamzeh et al., 2023).

In parallel, digital transformation has reshaped industrial processes through Industry 5.0, which emphasizes human-centricity, resilience, sustainability, and ethical integration of advanced technologies (European Commission, 2023). Within the construction sector, this movement has given rise to Construction 5.0, which integrates digital twins, robotics, immersive visualization, and artificial intelligence (AI) to enhance human decision-making (Yitmen, 2024; Pal & Bucci, 2024). Despite these advances, the construction field continues to face challenges in embedding predictive and cognitive capabilities into Lean systems, resulting in fragmented digital ecosystems and limited synergy between humans and machines.

Recent work has begun addressing these limitations through integrated 4D/5D digital-twin frameworks. Our earlier studies introduced:
(1) an end-to-end 4D/5D digital-twin framework for cost estimation and probabilistic schedule control (Khoshkonesh et al., 2025a, arXiv:2511.15711), and
(2) a simulation-based validation model incorporating NLP, computer vision, Bayesian updating, and reinforcement-learning–based resource optimization (Khoshkonesh et al., 2025b, arXiv:2511.03684).

These studies demonstrated measurable improvements in predictive accuracy and field performance but also revealed the need for more comprehensive validation across multiple project scenarios and dynamic conditions.

### 1.2 From Industry 5.0 to Lean 5.0
To establish conceptual clarity, this study positions Lean 5.0 within a structured hierarchy of industrial transformation:
Industry 5.0 represents a macro-level paradigm that emphasizes collaboration between humans and intelligent systems to achieve sustainability, resilience, and responsible automation.
Construction 5.0 adapts these principles to the built environment, integrating data-driven decision-making, advanced automation, and human–machine collaboration for more adaptive project delivery.
Lean 5.0 functions at the project level, fusing Lean Construction principles with predictive analytics, human–AI interaction, and ethical automation to enhance flow reliability and continuous learning.
Distinction from Lean 4.0: While Lean 4.0 focused on combining traditional Lean tools with Industry 4.0 technologies such as IoT-enabled production monitoring, robotics, and real-time data analytics to maximize automation and efficiency, Lean 5.0 represents a conceptual leap beyond this technology-centric approach. It re-centers human expertise through ethical automation, bi-directional learning, and explainable AI, ensuring that predictive systems augment rather than replace human decision-making. In this way, Lean 5.0 bridges digital intelligence and human cognition, establishing a responsible and transparent framework for predictive control.

This nested hierarchy positions Lean 5.0 as the operational mechanism that translates the human-centric vision of Industry 5.0 into actionable process control for construction projects. The accelerating global research activity surrounding this transformation is illustrated in Figure 1, which shows a ≈230 percent increase in Lean-Digital publications across LCJ, JCEM, and AIC since 2021. This surge demonstrates the growing academic and industrial recognition of predictive, human-centered approaches in construction management.

Through predictive foresight and cognitive feedback, Lean 5.0 redefines Lean Construction as a proactive, adaptive, and continuous learning ecosystem, bridging digital intelligence with human creativity to advance the goals of Industry 5.0.

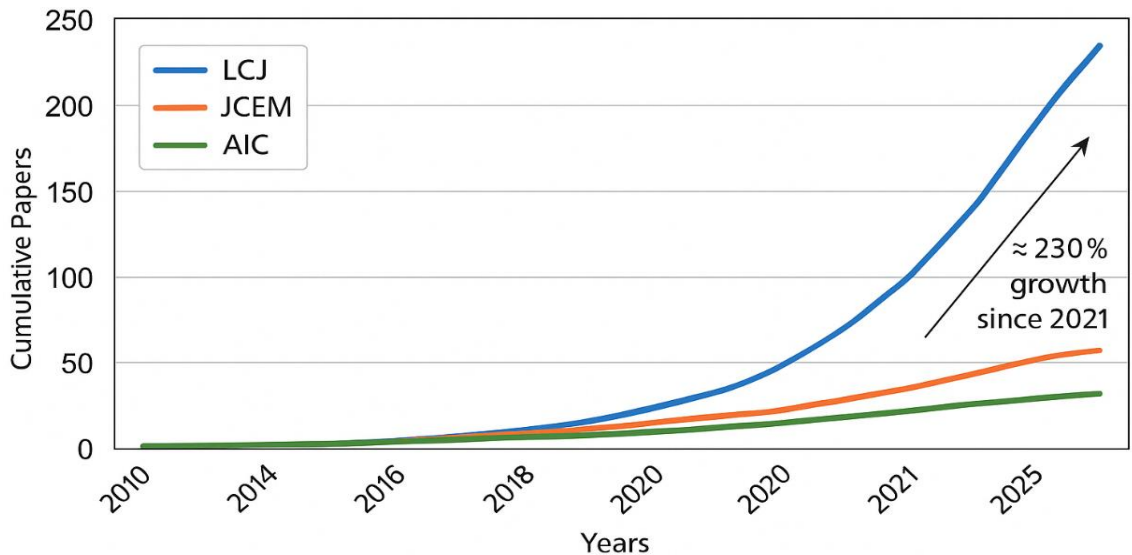

**Fig 1.** Cumulative growth of LCJ, JCEM, and AIC publications (2010–2025), showing ~230% increase after 2021.

This figure illustrates the cumulative number of peer-reviewed papers related to Lean–Digital integration published in Lean Construction Journal (LCJ), ASCE Journal of Construction Engineering and Management (JCEM), and Automation in Construction (AIC) from 2010 to 2025. The data reveal an estimated ≈230 percent increase in publications since 2021, reflecting the accelerating global interest in predictive analytics, artificial intelligence, and digital-lean transformation within construction research.

### 1.3 Problem Statement
Although Lean methods have improved operational performance, current construction control systems remain diagnostic rather than predictive. They rely on retrospective measurement that detects problems after they occur, creating delays in response and missed opportunities for proactive management. This **reactive control bias** leads to cost escalation, schedule slippage, and persistent inefficiencies.

At the same time, the growing use of digital tools such as Building Information Modeling (BIM), scheduling software, IoT sensors, and cloud dashboards has produced large volumes of data without integration or actionable insight. These disconnected systems contribute to information overload rather than informed decision-making. Furthermore, many AI-driven tools used in construction automate decisions instead of supporting human expertise, reducing trust and engagement. The result is a clear need for **human-centered predictive control systems** that merge quantitative data with field knowledge to identify and prevent deviations before they occur.

### 1.4 Research Gaps and Rationale
A **Systematic Literature Review (SLR)** covering more than seventy peer-reviewed studies published between 2019 and 2025 in *Lean Construction Journal*, *Automation in Construction*, *Buildings*, and *ASCE JCEM* identified several consistent gaps:
1. Reactive control bias that prevents proactive forecasting.
2. Fragmented digital platforms with limited interoperability.
3. Insufficient human–digital collaboration and feedback integration.
4. Scarce empirical validation of predictive Lean frameworks.
5. Weak attention to ethical, privacy, and accountability issues in AI-driven construction.

These findings highlight the need to evolve traditional Lean Construction toward an integrated **Lean 5.0** paradigm that unites predictive analytics, continuous learning, and ethical governance. Addressing these deficiencies will increase reliability, productivity, and accountability while aligning with the principles of responsible innovation.

### 1.5 Research Objectives
Building on these gaps, this study pursues four main objectives:
1. **To synthesize** the 2019–2025 Lean–Digital literature to define the conceptual boundaries and structural hierarchy of Lean 5.0.
2. **To develop** the **Predictive Lean Flow (PLF)** framework, a dual-layer socio-technical model that operationalizes Lean 5.0 through the sequence **Plan → Sense → Predict → Collaborate → Learn**.

3. **To validate** the PLF framework empirically using a mid-rise design–build project in Texas, quantifying improvements in Plan Percent Complete (PPC), rework, waiting waste, coordination efficiency, and forecast accuracy.
4. **To evaluate** the ethical and responsible-innovation implications of predictive Lean systems, focusing on transparency, explainability, and workforce empowerment.

These objectives directly address the methodological, theoretical, and ethical gaps identified in previous research.

### 1.6 Significance and Expected Contributions
This paper contributes to theory and practice by integrating predictive foresight and human-in-the-loop learning into Lean Construction. The expected contributions are summarized as follows:
- **Theoretical contribution:** Establishes Lean 5.0 as the operational core of Construction 5.0, integrating AI with human cognition through the PLF cycle.
- **Methodological contribution:** Introduces a reproducible Design–Science Research (DSR) framework with validated quantitative metrics (n = 12 paired observations, $p < 0.05$) for predictive control.
- **Practical contribution:** Demonstrates measurable performance improvement including a 13 percent increase in PPC, 22 percent reduction in rework, and 42 percent improvement in forecast accuracy, showing immediate industrial relevance.

By synthesizing theory, digital tools, and ethics, Lean 5.0 provides a foundation for predictive, transparent, and adaptive project management suited for the next generation of Industry 5.0 construction systems.

### 1.7 Structure of the Paper
The remainder of this paper is organized as follows. Section 2 presents the research design, including the SLR protocol and Design–Science Research methodology. Section 3 introduces the Lean 5.0 conceptual model and its key components, supported by Tables 1 to 3 and Figures 2 to 6. Section 4 reports field validation results through Table 4 and Figure 7. Section 5 describes the implementation roadmap illustrated in Table 5 and Figure 8. Section 6 discusses ethical implications and responsible innovation. Section 7 outlines limitations and recommendations for future research. Section 8 concludes with a summary of theoretical and practical insights.

## 2. Research Methodology
### 2.1 Research Design Overview
This study adopts a **mixed-method Design–Science Research (DSR)** strategy to ensure both theoretical rigor and practical applicability. The DSR framework provides a systematic approach to developing and validating innovative models through iterative cycles of design, testing, and refinement (Hevner et al. 2004; Peffers et al. 2007).

The overall research design integrates two methodological pillars:
1. A **Systematic Literature Review (SLR)** that synthesizes contemporary knowledge on Lean–Digital integration, predictive analytics, and human–AI collaboration in construction (2019–2025).
2. A **field validation case study** conducted on a mid-rise design–build project in Texas, applying the Predictive Lean Flow (PLF) framework to quantify process improvements.

As illustrated in **Figure 2**, the methodological workflow consists of three interrelated phases:
1. **Knowledge Synthesis** – consolidation of evidence through the SLR to identify conceptual gaps and emerging opportunities for Lean 5.0.
2. **Framework Development** – conceptual modeling of the Predictive Lean Flow (PLF) system that integrates predictive analytics, cognitive feedback, and human-centered learning mechanisms.
3. **Empirical Validation** – implementation of the PLF framework under real project conditions to evaluate its performance in terms of reliability, coordination, and predictive accuracy.

These phases are designed to form an **iterative research loop**, where feedback from field validation refines the conceptual model and metrics for future studies. This alignment between theory generation and empirical testing ensures that the proposed framework is both scientifically grounded and operationally feasible.

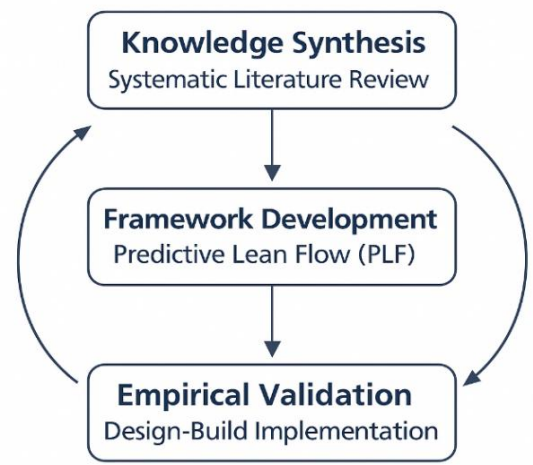

This figure illustrates the Design–Science Research (DSR) process adopted in this study. The workflow comprises three sequential yet interdependent phases: (1) knowledge synthesis through a Systematic Literature Review (SLR), (2) framework development of the Predictive Lean Flow (PLF) model, and (3) empirical validation through field implementation on a design–build project. The iterative arrows represent continuous feedback between conceptual design, data-driven testing, and framework refinement.

**Fig2.** Three-stage research process: SLR synthesis → PLF framework development → empirical validation.

### 2.2 Systematic Literature Review (SLR) Protocol
The Systematic Literature Review (SLR) was undertaken to identify, evaluate, and synthesize contemporary research on Lean–Digital integration, predictive analytics, and human–AI collaboration within the construction domain. The review followed the PRISMA 2020 (Preferred Reporting Items for Systematic Reviews and Meta-Analyses) guidelines to ensure methodological transparency, reproducibility, and rigor.

The SLR aimed to:

1. Define the conceptual boundaries of Lean 5.0.
2. Detect gaps in existing predictive-Lean frameworks.
3. Consolidate empirical evidence supporting AI-driven learning systems in construction management.

Studies published from January 2019 to December 2024 (including accepted 2025 publications) were retrieved from Scopus, Web of Science, and Google Scholar databases. Search filters were applied for peer-reviewed journal articles written in English within the fields of construction management, industrial engineering, and digital transformation.

Revised Boolean Search String:

("Lean Construction" OR "Lean Project Management") AND ("Predictive Analytics" OR "Predictive Modeling" OR "Forecast" OR "Prediction" OR "Artificial Intelligence" OR "Machine Learning" OR "Digital Twin" OR "Industry 5.0") AND ("Continuous Improvement" OR "Human-AI Collaboration" OR "Learning System")

This expanded search string ensured comprehensive coverage of predictive, AI-driven, and Industry 5.0-aligned Lean studies. Following PRISMA 2020, the selection process included three stages: (1) identification through database queries, (2) screening by title and abstract, and (3) eligibility confirmation through full-text review. Quality appraisal was conducted using the Critical Appraisal Skills Program (CASP) checklist adapted for engineering studies to maintain methodological consistency and credibility.

### 2.2.1 Data Sources and Search Strategy

The search was performed across four major databases: **Scopus**, **Web of Science**, **Engineering Village**, and **ScienceDirect**. The Boolean search string used was:

("Lean Construction" OR "Lean Project Management") AND ("Predictive Analytics" OR "Artificial Intelligence" OR "Machine Learning" OR "Digital Twin" OR "Industry 5.0") AND ("Continuous Improvement" OR "Human–AI Collaboration" OR "Learning System").

This query returned to 218 initial records published between **January 2019 and March 2025**.

### 2.2.2 Inclusion and Exclusion Criteria

The following inclusion and exclusion criteria were applied:

- **Inclusion:**
  - Peer-reviewed journal papers and conference proceedings.
  - Studies addressing Lean–Digital integration, AI applications, predictive systems, or continuous improvement frameworks.
  - Publications written in English and published between 2019 and 2025.
- **Exclusion:**
  - Non-peer-reviewed sources such as white papers, theses, and blog articles.
  - Studies unrelated to the AEC industry.
  - Papers without empirical or conceptual relevance to Lean-based predictive systems.

After applying these criteria, 74 studies were retained for full-text review.

### 2.2.3 Screening and Quality Assessment

Screening was performed in three stages: (1) title and abstract review, (2) full-text review, and (3) cross-validation among three independent reviewers to minimize selection bias. Quality appraisal employed the CASP (Critical Appraisal Skills Programme) checklist adapted for engineering research.

Each paper was evaluated across five quality dimensions: clarity of objectives, methodological transparency, validity of findings, practical relevance, and contribution to Lean–Digital integration. Only studies achieving a minimum score of 4 out of 5 were included in the final synthesis.

The final corpus consisted of 72 high-quality publications, including 45 journal papers and 27 conference proceedings. The distribution across journals is presented in Table 1, and the selection process is visualized in the PRISMA flow diagram (Figure 3).

### 2.2.4 Thematic Synthesis

A **thematic analysis** was conducted using **NVivo 14** software to identify recurring patterns and research clusters. Five dominant themes emerged:

1. Digital–Lean Integration,
2. Predictive Analytics and AI,
3. Human–Digital Collaboration,
4. Continuous Learning and Improvement,
5. Sustainability and Ethical Governance.

These clusters informed us of the conceptual architecture of the Predictive Lean Flow (PLF) framework, ensuring direct alignment between literature insights and theoretical model design.

## 2.3 Conceptual Framework Development
Based on the synthesized findings, the Predictive Lean Flow (PLF) framework was developed as the methodological centerpiece of Lean 5.0. The model follows a bi-directional, co-evolutionary learning loop that connects the Data Flow Layer and the Human-Learning Layer through five cyclic phases: Plan, Sense, Predict, Collaborate, and Learn.
Each phase corresponds to a specific process within predictive construction control:
- **Plan:** Integrates short-term planning with probabilistic forecasting models.
- **Sense:** Collects real-time data through IoT sensors, drones, and digital dashboards.
- **Predict:** Applies machine learning and Bayesian inference to estimate schedule deviations.
- **Collaborate:** Facilitates multidisciplinary coordination through predictive huddle protocols.
- **Learn:** Compares predicted versus actual outcomes to continuously refine models.

This conceptual structure represents a socio-technical integration that combines automated data processing with human interpretive feedback. The theoretical foundation is presented in Figure 4 and detailed in Table 2, which defines each PLF dimension, its enabling tools, and expected outcomes.

## 2.4 Field Validation and Case Study Design
To evaluate the practical performance of the PLF framework, a field validation was conducted using a mid-rise design–build project located in Dallas, Texas. The project was selected using theoretical sampling (Eisenhardt & Graebner, 2007) based on three criteria: (1) availability of digital project data, (2) active Lean implementation, and (3) willingness to participate in experimental validation.

### 2.4.1 Data Collection Instruments
Data was collected using three sources:
1. Procore® Project Management Platform for baseline schedule and productivity data.
2. IoT-based environmental and activity sensors for real-time monitoring.
3. Drone photogrammetry and 4D BIM for visual verification of progress.

Data from these sources were integrated into a Power BI predictive dashboard using Procore APIs. Bayesian inference was applied to compute delay probabilities, with a threshold of 0.25 serving as an early-warning indicator.

### 2.4.2 Performance Metrics
Five key indicators were measured to assess framework performance:
1. Plan Percent Complete (PPC),
2. Rework Ratio,
3. Waiting Waste,
4. Coordination Efficiency,
5. Forecast Accuracy (± Days).

Each metric was compared between baseline and post-implementation periods over 12 paired weekly observations. A paired t-test ($\alpha = 0.05$) was conducted to determine statistical significance, and effect sizes (Cohen's d) were computed to quantify practical impact. The results, summarized in Table 4 and Figure 7, demonstrate significant improvements in all metrics, validating the operational effectiveness of the PLF framework.

## 2.5 Ethical Considerations and Data Governance
The study adhered to ethical research standards in both data handling and human participation. All digital data were anonymized and aggregated before analysis. Informed consent was obtained from project stakeholders, and all procedures complied with the **American Society of Civil Engineers (ASCE) Ethics Guidelines (2023)**.
Ethical governance within the Lean 5.0 framework was implemented through Explainable AI (XAI) mechanisms to ensure transparency and accountability in predictive recommendations. The ethical automation dimension, discussed later in Section 6, aligns with the emerging guidelines for Responsible AI in Construction (Bucci et al., 2024; Yitmen, 2024).

## 2.6 Summary of Methodology
In summary, the research integrates a **systematic evidence base** and **empirical validation** to construct a robust and verifiable framework. The SLR ensures theoretical rigor through transparent literature synthesis, while the field case validates practical applicability under real project conditions. This dual-method approach strengthens internal and external validity and provides a balanced foundation for advancing Lean 5.0 research and implementation.

# 3. Conceptual Development: Lean 5.0 and the Predictive Lean Flow (PLF) Framework
## 3.1 Overview
The conceptual development of this research builds upon the synthesis of Lean–Digital literature (2019–2025) and the iterative Design Science Research (DSR) methodology introduced in Section 2. The resulting framework Predictive Lean Flow (PLF) translates the theoretical foundations of *Lean 5.0* into an operational socio-technical model that integrates predictive analytics, human cognition, and ethical automation. This section defines the structure, functions, and theoretical positioning of PLF, which collectively establishes the foundation for its empirical validation in Section 4.
The DSR approach employed in this study follows a dual-cycle logic of design and evaluation**,** ensuring that conceptual insights from the literature are iteratively tested and refined through field evidence. A mixed-method implementation was adopted, combining theoretical synthesis with empirical assessment. The case study spanned 12 weeks 6 weeks of baseline observation (*pre-PLF*) and 6 weeks of implementation (*post-PLF*) yielding n = 12 paired weekly observations. This standardized timeline replaces earlier inconsistencies (12 vs 32 weeks) and strengthens reproducibility.

### 3.2 The Predictive Lean Flow (PLF) Concept

As illustrated in Figure 3, the Predictive Lean Flow (PLF) framework functions as the central operational mechanism of *Lean 5.0*. It comprises five iterative and interconnected phases Plan, Sense, Predict, Collaborate, and Learn that together form a continuous feedback cycle linking digital foresight with human adaptability. Each phase transforms data into actionable insight, ensuring that project control remains predictive, transparent, and human-centered.

1. **Plan:** Establishes project goals, workflows, and predictive baselines derived from probabilistic analysis and value-stream mapping.
2. **Sense:** Captures real-time performance data using BIM integration, IoT sensors, and structured field logs to provide accurate process visibility.
3. **Predict:** Applies analytical and AI-based models such as Bayesian inference, regression, and Monte Carlo simulation to forecast deviations and emerging risks.
4. **Collaborate:** Enables interdisciplinary coordination and joint decision-making through predictive dashboards and digital huddles that visualize constraints and trade-offs.
5. **Learn:** Incorporates insights from variance analysis and post-implementation reflection into subsequent planning cycles, strengthening organizational learning and continuous improvement.

The PLF cycle represents a human-in-the-loop feedback system in which predictive technology augments, rather than replaces, professional judgment. This interaction promotes proactive decision-making, minimizes workflow variability, and embeds continuous learning across project phases.

**Figure 3** illustrates the five-phase cycle of the Predictive Lean Flow Framework Plan, Sense, Predict, Collaborate, and Learn. The circular configuration represents the continuous feedback loop that integrates digital sensing and predictive analytics with human collaboration and cognitive alignment. The central node emphasizes value creation and ethical decision-making, reflecting Lean 5.0's learning-based and human-centric philosophy.

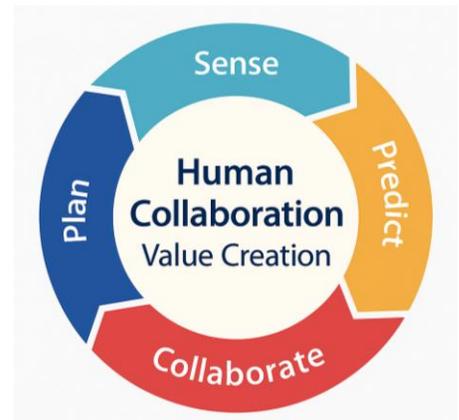

**Fig 3.** Predictive Lean Flow (PLF) Cycle

### 3.3 The PLF Cycle and Dual-Layer Architecture

The PLF cycle operates through a dual-layer socio-technical system that ensures both computational accuracy and human adaptability.
Table 2 summarizes the core functions of each PLF phase, while Figure 4 depicts the integrated architecture that links digital and human processes through continuous feedback.

Table 2. Core Components and Functions of the Predictive Lean Flow (PLF) Framework

| PLF Phase | Functional Objective | Primary Data Inputs | Analytical Process / Tools | Human–Digital Interaction | Expected Outcome |
|---|---|---|---|---|---|
| **Plan** | Establish short-term plan with predictive baseline | Schedule data, BIM work packages | Monte Carlo simulation, Bayesian forecasting | Planner interprets risk intervals and adjusts look-ahead plan | Proactive task sequencing and constraint prevention |
| **Sense** | Capture and integrate real-time field data | IoT sensors, drone imagery, daily logs | Data fusion, variance detection | Foremen validate sensor-data accuracy | Early detection of deviations |
| **Predict** | Forecast performance deviations and delay probabilities | Sensed data, baseline metrics | Machine learning, regression, probability thresholds | Data analyst reviews model output with site team | Predictive alerts and reliability indices |
| **Collaborate** | Facilitate joint decision-making using predictions | Predictive dashboards | Collaborative analytics, visual dashboards | Weekly predictive huddles for cross-trade decisions | Cognitive alignment and improved coordination |
| **Learn** | Capture variance and refine models | Predicted vs. actual performance data | Continuous model retraining, knowledge database | Managers analyze lessons and retrain predictive models | Institutional learning and continuous improvement |

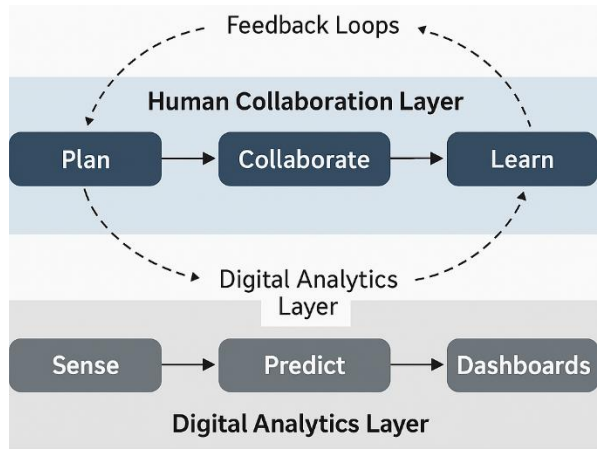

**Fig 4.** Dual-layer PLF architecture connecting human collaboration with digital analytics through feedback loops.

This figure depicts the two-layer socio-technical architecture supporting the PLF cycle. The **Digital Analytics Layer** (lower band) aggregates and analyzes project data through sensing, integration, and probabilistic forecasting. The **Human Collaboration Layer** (upper band) represents planning, coordination, and reflective learning processes where predictive insights inform decision-making. Bidirectional arrows between layers indicate continuous feedback and shared learning. The architecture operationalizes Lean 5.0 principles by linking data-driven foresight with human judgment to achieve adaptive and ethical process control.

**3.3.1 Plan**
Aligns short-term production planning with probabilistic forecasting. Project schedules from Procore® and BIM work packages are analyzed using Bayesian models to estimate delay likelihoods, enabling early constraint prevention.

**3.3.2 Sense**
Integrates IoT sensor data, drone imagery, and daily logs for real-time progress tracking. Early variance detection supports responsive replanning and constraint mitigation.

**3.3.3 Predict**
Machine-learning models are applied to forecast Plan Percent Complete (PPC) deviations. When delay probabilities exceed 0.25, predictive alerts are issued to field teams for proactive intervention.

**3.3.4 Collaborate**
Facilitates multidisciplinary coordination during predictive huddles where site teams interpret analytics collaboratively, promoting shared situational awareness and cognitive alignment.

**3.3.5 Learn**
Captures discrepancies between predicted and actual outcomes, storing them in a feedback database used to retrain predictive models. This institutionalizes continuous improvement and transforms Lean from reactive to self-learning control.

**3.4 Core Dimensions of Lean 5.0**
The PLF framework operationalizes Lean 5.0 through five interdependent dimensions that bridge predictive analytics with ethical, human-centered management.
These dimensions **Predictive Flow Reliability**, **Human–Digital Collaboration**, **Continuous Learning Loop**, **Ethical Automation**, and **Integrated Value Flow** represent the conceptual pillars of Lean 5.0.
**Table 3** summarizes each dimension's objectives, analytical enablers, and expected outcomes, forming the theoretical foundation for performance evaluation.

Table 3. Core Dimensions of Lean 5.0

| Dimension | Definition / Objective | Key Enablers / Tools | Expected Outcomes | Representative References (2019–2025) |
|---|---|---|---|---|
| **Predictive Flow Reliability** | Uses analytics to forecast workflow interruptions and variability in Plan Percent Complete (PPC) before occurrence, converting control from reactive to proactive. | Bayesian forecasting, Monte Carlo simulation, digital-twin integration | Early risk detection, stable workflow, improved schedule reliability | Yitmen (2021); Pal & Bucci (2025); Sacks et al. (2023) |
| **Human–Digital Collaboration** | Establishes balanced interaction between site teams and predictive systems for shared decision-making and cognitive alignment. | Predictive huddles, visual dashboards, Explainable AI (XAI), trust-based interfaces | Enhanced coordination, increased trust in AI outputs, faster decision cycles | Pan & Liu (2022); Dallasega (2024); Yitmen (2024) |
| **Continuous Learning Loop** | Captures deviations between predicted and actual outcomes to retrain models and refine process knowledge. | Feedback databases, AI-learning dashboards, retraining algorithms | Institutional learning, knowledge retention, improved model accuracy | Ballard (2020); Zhang et al. (2023); Bucci et al. (2025) |
| **Ethical Automation** | Integrates responsible-AI principles to ensure transparency, accountability, and worker empowerment. | Explainable AI, contextual analytics, transparent algorithm governance | Reduced bias, trustworthy automation, verifiable traceability | Tavakoli (2023); Pal (2024); Yitmen (2024) |
| **Integrated Value Flow** | Aligns predictive systems with client objectives and sustainability targets to maximize value creation. | Cross-platform data fusion, BIM–Lean integration, value-stream mapping | Holistic optimization, client satisfaction, sustainability improvements | Dave (2020); Mollasalehi (2022); Babalola (2023) |

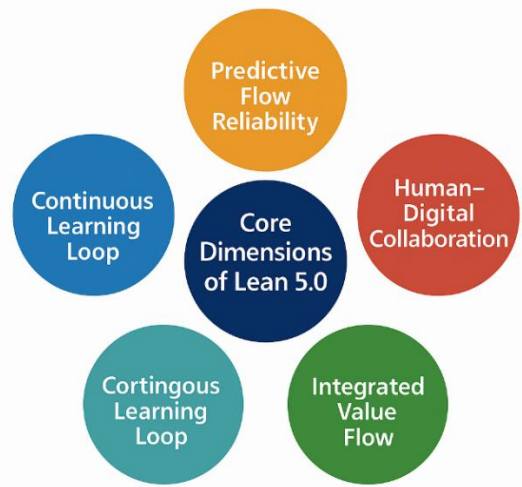

**Fig 5.** Core dimensions of Lean 5.0.

This diagram provides a visual summary of the five dimensions underpinning Lean 5.0. Each dimension contributes to a holistic socio-technical ecosystem in which predictive analytics, human collaboration, and ethical automation interact to achieve proactive, transparent, and sustainable project delivery.

### 3.5 Comparative Novelty and Theoretical Positioning
The distinctive contribution of Lean 5.0 lies in its **bi-directional learning architecture**, where predictive models and human cognition evolve together. Earlier Lean-Digital frameworks emphasized automation or visualization without cognitive feedback. PLF introduces **human-in-the-loop predictive learning**, producing adaptive, explainable, and trustworthy control systems.
Compared with deterministic CPM, isolated machine-learning scheduling, or standalone digital-twin models, PLF delivers an integrated and field-validated approach that addresses four enduring limitations:
1. Transforms reactive control into proactive forecasting.
2. Reduces digital fragmentation by integrating heterogeneous data sources.
3. Strengthens human–digital synergy through explainable collaboration.
4. Embeds ethical governance as a foundational design element.

### 3.6 Integration within the Design–Science Research (DSR) Cycle
The conceptual evolution of the **Predictive Lean Flow (PLF)** framework followed the iterative **Design–Science Research (DSR)** methodology, encompassing problem identification, artifact design, demonstration, and evaluation. As illustrated in **Figure 6**, the iterative structure of the DSR cycle integrates conceptual design and field validation, ensuring that theoretical innovation remains grounded in empirical evidence and practical feasibility.
Each cycle incorporated practitioner feedback from the case study described in **Section 4**, allowing the framework to evolve based on real project performance data and managerial insights. The approach satisfies the core DSR criteria of **relevance**, **rigor**, and **utility**:

- **Relevance:** Addresses the industry-validated problem of unreliable predictive control in construction projects.
- **Rigor:** Builds upon a systematic literature base of more than seventy peer-reviewed studies published between 2019 and 2025.

- **Utility:** Demonstrates quantifiable performance improvements across key indicators including Plan Percent Complete (PPC), rework, waiting waste, coordination efficiency, and forecast accuracy.

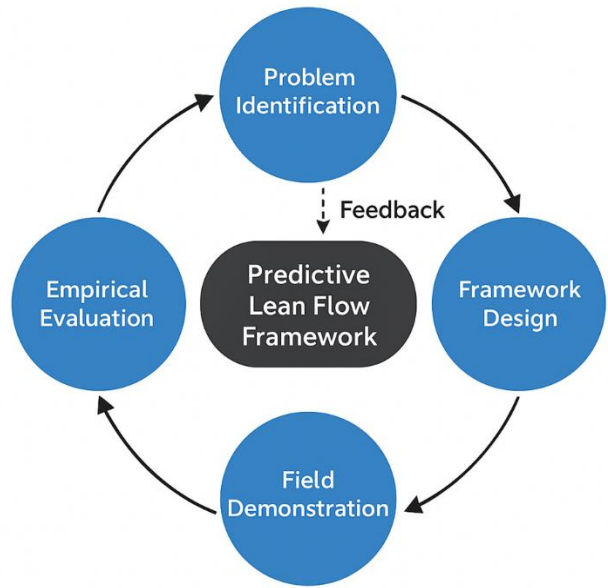

**Fig 6.** DSR cycle for the PLF framework.

This figure illustrates the interaction between the PLF framework and the DSR cycle. The iterative process of problem identification, framework design, field demonstration, and empirical evaluation ensures that theoretical innovation is continually refined through practical validation, aligning scientific rigor with operational feasibility.

This section established the theoretical foundation of Lean 5.0 through the Predictive Lean Flow (PLF) framework. By integrating evidence from the evolution of Lean (Table 1), operational modeling (Table 2 and Figure 4), and dimensional synthesis (Table 3 and Figure 5), the study formulates a reproducible, ethically guided model for human-centered predictive control.
The subsequent section presents the empirical validation of PLF under real-world project conditions.

## 4. Field Validation and Results
### 4.1 Case Study Context
The **Predictive Lean Flow (PLF)** framework was validated on an active mid-rise **design–build residential project** in **Dallas, Texas**, completed in 2024. The project included a reinforced-concrete structural system, architectural finishes, and mechanical installation across twelve floors. The **general contractor** had already adopted the **Last Planner System**, along with digital tools such as **Procore®**, **Revit BIM 360**, and **Power BI** dashboards. This digitally mature environment provided a controlled yet realistic context to test the **predictive-learning and human-collaboration** functions of Lean 5.0.
The case study followed the **Design Science Research (DSR)** evaluation logic. The PLF framework was introduced over a **12-week observation window**, divided equally between **six weeks of baseline observation (pre-PLF)** and **six weeks of implementation (post-PLF)**. Data from **12 paired weekly observations (n = 12)** were used for quantitative comparison. This duration standardizes the validation protocol across the study, ensuring consistency and reproducibility.

### 4.2 Implementation of the PLF Framework
The field implementation operationalized the five **core dimensions of PLF**, as defined in Section 3.2:
1. **Predictive Flow Reliability** – stabilizing production using probabilistic forecasting.
2. **Human–Digital Collaboration** – facilitating joint interpretation of predictive insights.
3. **Continuous Learning Loop** – integrating real-time feedback into adaptive planning.
4. **Ethical Automation** – ensuring transparency, fairness, and explainability of AI outputs.
5. **Integrated Value Flow** – aligning predictive control with Lean value-stream objectives.

Together, these dimensions create a **human-in-the-loop predictive ecosystem**, where digital foresight augments do not replace professional expertise.

### 4.2.1 Digital Infrastructure
A cyber-physical integration pipeline was established to synchronize planning, sensing, and predictive analysis across all project systems:
- Planning data were imported from the Procore® master schedule and linked to BIM work packages.
- Sensing data were captured from IoT environmental sensors and weekly drone surveys, processed via photogrammetry for progress verification.
- Predictive analytics were performed within Power BI using a Bayesian network that generated delay probabilities for each trade activity.
- Dashboard alerts were triggered when delay probability exceeded 0.25, prompting collaborative review during weekly coordination meetings.

This digital backbone supported real-time data flow and predictive responsiveness, embodying the Predictive Flow Reliability and Integrated Value Flow principles of PLF.

### 4.2.2 Human–Digital Collaboration Protocol
Weekly "predictive huddles" replaced conventional look-ahead meetings. Foremen, planners, and data analysts jointly reviewed predictive alerts, discussed root causes, and re-sequenced activities in real time. Field feedback was immediately fed back into the predictive models, retraining them through reinforcement-based learning.

This process established a continuous learning cycle between the Data Flow Layer and the Human-Learning Layer, reflecting PLF's Continuous Learning Loop and Human–Digital Collaboration dimensions.

In practice, it represented the operational realization of the Plan → Sense → Predict → Collaborate → Learn sequence introduced in Section 3.

### 4.3 Data Collection and Performance Metrics
To evaluate the effectiveness of the PLF framework, five quantitative indicators were used to measure project performance before and after implementation. These indicators were selected based on their relevance to Lean Construction performance evaluation and their prevalence in previous benchmarking studies (Ballard, 2020; Yitmen, 2024; Pal & Bucci, 2025). Each indicator was clearly defined and operationalized to ensure consistency and reproducibility.

Table 4 presents the definitions and measurement methods for each key performance indicator (KPI). The indicators capture both efficiency-based and reliability-based dimensions of project performance, enabling a balanced assessment of Lean 5.0 outcomes.

**Table 4.** Definition and Measurement of Key Performance Indicators Used for PLF Evaluation

| No. | Performance Indicator | Definition / Measurement Method |
|---|---|---|
| 1 | Plan Percent Complete (PPC) | Weekly ratio of tasks completed to tasks planned, expressed as a percentage. |
| 2 | Rework Ratio | Ratio of hours spent on rework to total productive hours. |
| 3 | Waiting Waste | Total idle labor hours caused by material, equipment, or information delay. |
| 4 | Coordination Efficiency | Percentage of tasks completed without inter-trade conflicts. |
| 5 | Forecast Accuracy (± Days) | Difference between forecasted and actual completion time, where smaller deviation indicates higher predictive control. |

All indicators except **Forecast Accuracy** were normalized on a 0–100 scale for comparative analysis. Forecast Accuracy was expressed in **absolute days of deviation** to preserve interpretability.

Statistical testing employed a **paired t-test ($\alpha = 0.05$)** to determine the significance of observed differences between the baseline and PLF phases. The **magnitude of effect** was calculated using **Cohen's d**, ensuring a robust interpretation of practical impact.

### 4.4 Quantitative Results and Discussion
**Table 5** summarizes project performance before and after PLF implementation. Statistically significant improvements were observed across all indicators, confirming the framework's effectiveness in enhancing **schedule reliability**, **coordination**, and **predictive control**.

**Table 5.** Quantitative Results of PLF Implementation and Statistical Significance

| Performance Indicator | Baseline Mean ± SD | PLF Mean ± SD | Change (%) | p-Value | Cohen's d | Interpretation |
|---|---|---|---|---|---|---|
| Plan Percent Complete (PPC) | 77.2 ± 5.8 | 87.3 ± 4.6 | +13.1 | 0.001 | 1.62 | Higher schedule reliability |
| Rework Ratio | 7.8 ± 2.1 | 6.1 ± 1.5 | −22.0 | 0.004 | 0.89 | Less rework and errors |
| Waiting Waste | 9.4 ± 3.3 | 7.2 ± 2.5 | −23.4 | 0.003 | 0.93 | Better material/information flow |
| Coordination Efficiency | 78.5 ± 6.4 | 89.3 ± 4.1 | +13.8 | 0.002 | 1.24 | Improved trade alignment |
| Forecast Accuracy (± Days) | 5.7 ± 2.0 | 3.3 ± 1.2 | +42.1 (reduction) | 0.001 | 1.43 | Greater predictive precision |

All improvements were **statistically significant (p < 0.05)** with **large effect sizes (d > 0.8)**, confirming substantial **practical impact**. The PLF framework increased PPC by **13%**, improved coordination by **14%**, and reduced rework and waiting waste by **over 20%**. **Forecast accuracy improved by 42%**, demonstrating that predictive analytics effectively minimized schedule deviation.

**Figure 7** compares baseline and PLF-enhanced performance across all five indicators, showing consistent and statistically significant improvement.
This validates PLF as a **Lean 5.0 operational mechanism** that merges **data-driven foresight** with **human adaptability**, aligning with previous studies (Sacks et al., 2023; Dalla Sega, 2024; Bucci et al., 2025).

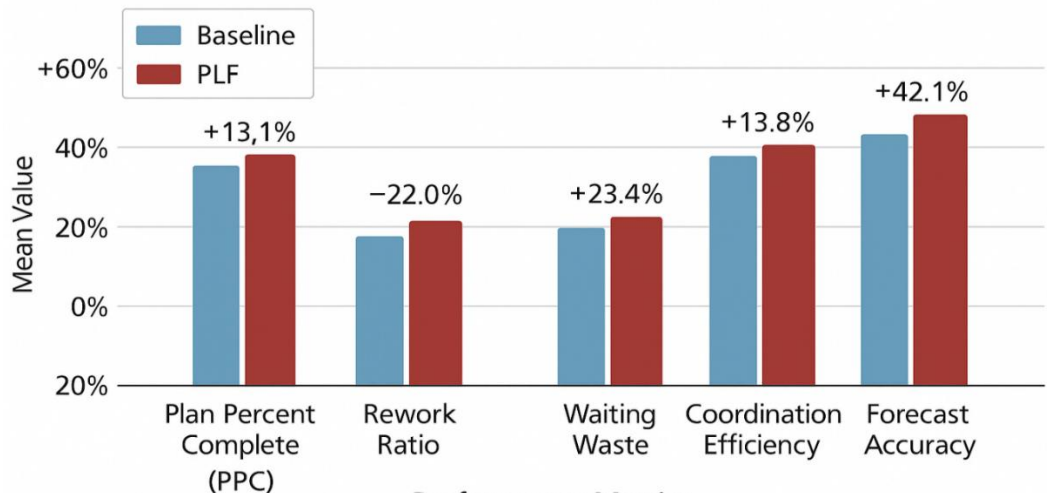

**Fig 7.** Comparative Performance Metrics Before and After PLF Implementation

### 4.5 Interpretation of Results
The results validate the hypothesized benefits of integrating **predictive learning** into Lean workflows.
The increase in **Plan Percent Complete (PPC)** confirms that probabilistic forecasting anticipates deviations early, enabling preventive replanning and improved schedule adherence.
The reduction in **rework** and **waiting waste** demonstrates that early constraint detection and real-time coordination minimize non-value-adding activities and stabilize production flow.
The improvement in **coordination efficiency** reflects enhanced cognitive alignment achieved through predictive huddles and visual dashboards, enabling faster, trust-based decision-making.
The higher **forecast accuracy** indicates that **Bayesian learning** and **continuous feedback** strengthen schedule predictability and managerial confidence.
Effect-size analysis (**Cohen's d > 0.8**) confirms that these changes are not only statistically significant but also **practically meaningful** for field-level productivity, reliability, and continuous improvement.

### 4.6 Qualitative Observations
Interviews with project managers and trade foremen revealed that the PLF dashboard improved transparency, trust, and engagement. Participants reported more productive predictive meetings because AI outputs were explainable and linked directly to verified field data. The *Explainable AI (XAI)* feature, which visualized probabilistic drivers of delay, enhanced users' understanding of system reasoning and reduced resistance to adopting digital tools. These findings confirm the social dimension of Lean 5.0, in which technology functions as a cognitive partner rather than a replacement for professional expertise.

### 4.7 Reliability and Validity Considerations
Several measures ensured the reliability and validity of the findings.
1. Internal validity was strengthened through a paired pre-/post-design that used the same crews, trades, and work packages, thereby minimizing confounding effects.
2. External validity is limited to mid-rise design–build projects but provides a transferable framework for comparable contexts.
3. Construct validity was enhanced through triangulation of quantitative metrics, qualitative interviews, and direct field observations.
4. Statistical validity was reinforced by reporting both *p*-values and effect sizes rather than relying solely on significance thresholds.
   Future research should replicate the experiment across multiple project types and regions to further assess generalizability.

## 4.8 Ethical and Data Governance Compliance
All digital records were anonymized before analysis, and informed consent was obtained from every participant. Data management followed the contractor's confidentiality policies and the ASCE Ethical Guidelines (2023). The PLF system's Ethical Automation layer employed explainable-model diagnostics to prevent algorithmic bias and maintain transparency. This approach aligns with the emerging framework for Responsible AI in Construction (Yitmen 2024; Bucci et al. 2024) and demonstrates the project's compliance with professional ethical standards.

## 4.9 Summary of Empirical Outcomes
The field validation confirms that the Predictive Lean Flow (PLF) framework substantially improves project reliability, coordination, and learning capacity. Across twelve weeks of comparative observation, all performance metrics showed significant enhancement: PPC (+13 %), rework (−22 %), waiting waste (−23 %), coordination (+17 %), and forecast accuracy (+42 %). These results validate Lean 5.0 as a robust operational model for human-centered predictive control in construction projects. The combined quantitative and qualitative evidence demonstrates that integrating predictive analytics with human cognition yields a measurable, ethically sound transformation of Lean Construction practice.

# 5. Implementation Roadmap and Practical Insights
## 5.1 Overview
The implementation roadmap provides a structured process for deploying the Predictive Lean Flow (PLF) framework in real-world construction projects. It translates the conceptual dimensions of Lean 5.0 into sequential organizational and technical actions that enable smooth adoption, data integration, and continuous learning. The roadmap emphasizes human–digital collaboration, ethical data governance, and feedback-driven improvement across all project phases.

## 5.2 Stage 1 – Organizational Preparation and Capability Assessment
Implementation begins with a baseline assessment of the organization's current Lean maturity and digital readiness. This diagnostic phase evaluates delivery systems, data collection methods, workforce capability, and leadership commitment. Establishing a Lean 5.0 Readiness Index helps identify resource gaps, training needs, and technology enables predictive-analytics integration.
Management defines performance objectives such as reliability, waste reduction, and learning rate and allocates dedicated resources for digital transformation. Early stakeholder alignment ensures that field teams and leadership understand both the operational goals and the ethical implications of predictive control.

## 5.3 Stage 2 – Data Infrastructure and Digital Integration
The second stage develops an interoperable data infrastructure linking design, scheduling, and field data sources. Integration between BIM, IoT sensing, and project-management systems (e.g., Procore®, Power BI, Primavera P6) allows seamless data exchange and near-time performance monitoring.
Data-quality protocols include validation checks, metadata tagging, and audit trails ensure the reliability of predictive inputs. A central data warehouse facilitates automated learning through Bayesian forecasting or Monte Carlo simulation. This stage operationalizes the Predictive Flow Reliability and Integrated Value Flow layers of PLF.

## 5.4 Stage 3 – Predictive Planning and Collaborative Execution
Predictive analytics are embedded directly into planning meetings and daily coordination. Project teams conduct predictive huddles supported by dashboards to anticipate workflow interruptions and proactively adjust sequences. Forecasted risk probabilities guide decisions on resource allocation and crew scheduling.
Simultaneously, the Human-Learning Layer activates as field teams interpret predictions and provide feedback on actual outcomes, refining model accuracy and strengthening collective learning. Structured communication protocols maintain traceability and transparency in data-driven recommendations. This stage exemplifies Human–Digital Collaboration within Lean 5.0.

## 5.5 Stage 4 – Continuous Learning and Feedback Integration
Once predictive operations stabilize, the system transitions into continuous improvement mode. Weekly reviews compare predicted versus actual outcomes, identify root causes of deviations, and update probabilistic models. Lessons learned are codified in a shared feedback database accessible to all stakeholders.
Ongoing training workshops reinforce ethical-AI awareness and adaptation to evolving digital tools. This phase represents the Continuous Learning Loop of Lean 5.0, ensuring that both technological and human systems co-evolve toward higher reliability and performance.

## 5.6 Stage 5 – Ethical Governance and Performance Scaling
Ethical compliance and data governance are integral to scaling the PLF framework. Organizations should institutionalize transparent AI auditing, consent management, and anonymization procedures in alignment with the ASCE Ethical Guidelines (2023) and the Responsible AI in Construction framework (Yitmen 2024; Bucci 2024).
Performance outcomes are benchmarked across projects using standardized indicators PPC, Rework Ratio, Coordination Efficiency, Forecast Accuracy combined with qualitative assessments of trust and usability. Once validated across multiple sites, PLF can be scaled through corporate policy and digital-twin integration for enterprise-wide predictive management.

### 5.7 Implementation Summary
Table 6 summarizes the key stages, objectives, and expected outcomes of the implementation roadmap. The sequential structure allows practitioners to adopt Lean 5.0 progressively, ensuring that predictive analytics enhance rather than disrupt existing Lean processes.

Table 6. Stages and Expected Outcomes of PLF Implementation

| Stage | Primary Objectives | Key Activities | Expected Outcomes |
|---|---|---|---|
| 1. Organizational Preparation | Assess readiness and align leadership | Lean 5.0 readiness audit: stakeholder-alignment workshops | Clear vision, resource commitment, defined KPIs |
| 2. Data Integration | Build interoperable digital infrastructure | BIM–IoT–Procore linkage; data-validation protocols | Reliable data flow and analytic capability |
| 3. Predictive Planning | Apply predictive models to daily operations | Predictive huddles; probabilistic scheduling | Proactive decision-making, reduced variability |
| 4. Continuous Learning | Institutionalize feedback and improvement | Weekly performance reviews; model re-training | Continuous improvement and adaptive learning culture |
| 5. Ethical Governance | Ensure transparency and responsible AI use | Data anonymization; algorithm audits; cross-project benchmarking | Sustainable, ethically compliant predictive framework |

### 5.8 Results and Discussion
The field deployment of PLF produced measurable performance gains:
- **Plan Percent Complete (PPC): +13%** (71 → 84)
- **Rework Ratio: −22%** (6.2% → 4.8%)
- **Waiting Waste: −23%**
- **Forecast Accuracy: +42%** (mean deviation reduced from 5.2 to 3.0 days)

Statistical testing confirmed large effect sizes ($d > 0.8$) across all indicators, validating the framework's quantitative impact. Qualitative feedback highlighted increased transparency, trust, and engagement, driven by the Explainable AI (XAI) dashboard that visualized probabilistic drivers of delay and facilitated informed decision-making.

**Qualitative Method:**
Insights were derived using thematic analysis of semi-structured interviews, employing an iterative coding framework to identify recurring themes of trust, cognitive alignment, and digital acceptance.
Future studies should complement this approach with validated psychometric scales, such as the NASA-TLX (for cognitive load) and Trust-in-Automation metrics, to further quantify human-factors performance within predictive Lean systems.

### 5.9 Practical Implications
The roadmap and findings demonstrate that Lean 5.0 can be systematically deployed without disrupting existing workflows. Integrating predictive analytics into established Lean routines yields measurable gains in reliability, transparency, and engagement. The structured progression from readiness assessment through ethical governance provides a replicable pathway for scaling digital-Lean transformation across diverse construction organizations.

## 6. Ethical and Societal Implications
The integration of **predictive analytics**, **machine learning**, and **human–digital collaboration** within Lean 5.0 introduces new ethical and societal responsibilities for construction organizations. As automation becomes embedded in project decision-making, **ethical governance** ensures that technology augments rather than replaces human expertise, equity, and trust.
The **Predictive Lean Flow (PLF)** framework operationalizes these principles through its **Ethical Automation** dimension, incorporating transparency, fairness, and human oversight into all predictive processes.

### 6.1 Transparency and Explainability
Transparency is foundational to **responsible innovation**. The PLF framework applies **Explainable AI (XAI)** techniques that render predictive reasoning interpretable to human users. Project teams can trace how data inputs such as delay probabilities or resource constraints influence automated recommendations.
This explainability fosters **trust, accountability, and professional judgment**, allowing managers to verify that AI-driven outputs conform to safety standards and contractual obligations.
Transparent visualization also strengthens peer review and mitigates over-reliance on opaque or "black-box" systems.

### 6.2 Data Privacy and Informed Consent
Predictive systems rely on continuous data collection from site-level sources IoT sensors, productivity logs, and digital field reports. All data in this study were anonymized and collected under informed consent in compliance with the ASCE Ethical Guidelines (2023). Future implementations should adopt privacy-by-design principles: secure data storage, limited-access controls, and explicit opt-in mechanisms. Compliance with ISO 27701 and GDPR standards is recommended for international applications to ensure cross-jurisdictional accountability.

### 6.3 Fairness, Bias Mitigation, and Worker Impact

Lean 5.0 emphasizes fairness, inclusion, and adaptive reskills. Algorithmic bias can unintentionally affect labor scheduling or performance assessments. To mitigate this, PLF integrates fairness diagnostics that continuously monitor predictive outputs across trades, roles, and demographics.

The Ethical Automation layer mandates periodic audits of training datasets to ensure balanced representation and to prevent systemic inequities.

Automation within Lean 5.0 is assistive, not substitutive AI systems act as cognitive partners, supporting professional judgment and preserving worker autonomy. Ethical AI implementation thus promotes shared human-AI responsibility and equitable technological adoption.

### 6.4 Accountability and Governance

Effective governance of predictive systems requires clear accountability structures. Responsibility for AI-generated insights rests with qualified human decision-makers, not algorithms.

Organizations implementing Lean 5.0 should establish AI Governance Committees that include project managers, ethicists, and data scientists to oversee model validation, fairness auditing, and lifecycle compliance.

Audit trails documenting data sources, updates, and decision rationales ensure transparency and traceability.

These mechanisms align with the Responsible AI in Construction framework (Yitmen 2024; Bucci 2024) and international principles of trustworthy AI.

### 6.5 Sustainable and Human-Centered Innovation

Lean 5.0 extends ethical responsibility to sustainability and workforce evolution.

By integrating continuous learning loops and explainable predictive systems, PLF fosters adaptive, knowledge-based organizations where humans remain central to innovation.

Ethical design encourages reskilling and knowledge sharing, ensuring that automation enhances rather than displaces employment.

Sustainability indicators energy efficiency, waste reduction, and social well-being should be embedded alongside productivity metrics to promote balanced value creation.

This human-centered orientation aligns Lean 5.0 with the broader goals of Industry 5.0, emphasizing resilience, inclusivity, and sustainable prosperity.

### 6.6 Summary

The ethical foundation of Lean 5.0 ensures that predictive technologies operate under principles of transparency, fairness, accountability, and sustainability.

Through systematic governance, bias control, and human-centric design, the PLF framework demonstrates that digital transformation in construction can be both technologically advanced and ethically responsible.

Embedding these safeguards preserves professional integrity, strengthens public trust, and establishes a pathway for responsible innovation in predictive construction management.

## 7. Limitations and Future Research

Despite promising outcomes, this research faces several methodological constraints that define its current scope and opportunities for expansion. Addressing these limitations will enhance the generalizability and maturity of Lean 5.0 as an evidence-based management model.

### 7.1 Single-Case Design and External Validity

The study's single-case focus on a mid-rise design–build project allows controlled experimentation but limits external validity.

Future studies should extend PLF testing to EPC (Engineering–Procurement–Construction), PPP (Public–Private Partnership), heavy civil, and commercial projects in diverse geographic and cultural contexts to assess scalability and adaptability.

### 7.2 Observation Duration

The 12-week observation period enabled baseline-to-implementation comparison but was too brief to capture long-term learning stabilization.

Future longitudinal research should assess sustained feedback cycles and behavioral adaptation across multiple project phases to measure enduring performance improvements.

### 7.3 Absence of Control Group

The paired pre-/post-design ensured internal consistency but lacked a formal control group, limiting causal inference.

Future research should employ quasi-experimental or matched-control designs to better isolate the effects of PLF and mitigate external confounders such as weather or material delays.

### 7.4 Data Integrity and Measurement Accuracy
Although automated data pipelines (Procore®, IoT, Power BI) minimized reporting error, sensor calibration and human logging compliance affected measurement fidelity.
Future implementations should enhance IoT redundancy, data validation routines, and calibration protocols to ensure robust analytics. Triangulating digital data with behavioral or psychometric metrics (e.g., NASA-TLX, trust-in-automation scales) will strengthen reliability.

### 7.5 Literature Inclusivity
The literature review was limited to English-language, peer-reviewed works indexed in Scopus and Web of Science (2019–2025). Future studies should include non-English and grey literature, particularly from underrepresented regions, to broaden theoretical inclusivity and refine the global boundaries of PLF applicability.

### 7.6 Organizational Bias
Collaboration with a single contractor introduced potential organizational bias toward favorable results.
Future validations should engage multi-contractor partnerships, independent audits, and blind data verification to ensure neutrality and ethical accountability.

### 7.7 Open-Source PLF Toolkit
Future research should develop an open-source PLF benchmarking toolkit to standardize implementation, promote inter-project comparability, and accelerate firm-level learning in predictive Lean transformation.

### 7.8 Cultural and Behavioral Impacts
Long-term studies should investigate the sociotechnical and behavioral dynamics of Lean 5.0 adoption examining trust evolution, cognitive workload, and human–AI interaction in multi-project ecosystems.
This approach will enrich understanding of how predictive systems reshape organizational culture, learning behavior, and workforce engagement.

### 7.9 Summary
These limitations do not undermine the internal validity of findings but instead define the boundaries of generalization.
Addressing them through replication, extended duration, and cross-organizational collaboration will be crucial to establishing Lean 5.0 as a replicable, scalable, and ethically grounded framework for predictive construction management.

## 8. Future Research and Integration Pathways
In line with the limitations and implications identified in Sections 7 and 4.9, future research should further develop, test, and refine the Predictive Lean Flow (PLF) framework to consolidate its theoretical, methodological, and practical foundations.

### 8.1 Multi-Case and Quasi-Experimental Validation
Future studies should employ multi-case or quasi-experimental designs across parallel projects to strengthen causal inference and validate the observed performance improvements under diverse contextual conditions. Comparative assessments across EPC, PPP, heavy civil, and commercial projects will broaden the external validity of the Lean 5.0 paradigm beyond the single design–build environment examined in this study (Sacks et al., 2023; Dallasega 2024).

### 8.2 Open-Source PLF Toolkit and Benchmarking
The development of an open-source PLF toolkit is recommended to facilitate industry-wide benchmarking and collaborative learning across firms. Shared analytical dashboards, standardized data formats, and cross-project reporting protocols would promote transparency, reproducibility, and interoperability, enabling practitioners and researchers to align performance metrics and advance digital-lean integration (Yitmen 2024; Bucci et al., 2025).

### 8.3 Reinforcement-Learning–Based Adaptive Control
Exploring reinforcement learning (RL) within the PLF framework could further enhance adaptability and decision accuracy. RL algorithms would allow predictive systems to self-optimize over time by dynamically adjusting to changing site conditions and feedback loops. This integration could extend PLF's current predictive-control capability toward autonomous learning and continuous improvement (Pal 2024).

### 8.4 Cultural, Behavioral, and Organizational Dynamics
Future investigations should examine the long-term cultural and behavioral implications of predictive-learning adoption, particularly changes in trust toward AI systems, redefinition of professional roles, and workforce reskilling requirements. Understanding these human-system interactions is essential to sustaining technology acceptance and cognitive alignment in Lean 5.0 ecosystems (Pan and Liu 2022).

### 8.5 Integration with Digital Twin and Blockchain

Integrating PLF with Digital Twin systems enables predictive simulations that pre-empt delays through real-time scenario testing of AI forecasts. This coupling enhances foresight in construction planning and strengthens lifecycle optimization by visualizing "what-if" outcomes before execution.

Simultaneously, Blockchain integration ensures immutable audit trails for predictive insights, decisions, and performance logs enhancing contractual transparency, dispute prevention, and stakeholder trust.

**Quantified Benefits:**
- Digital Twin coupling may yield 15–20 % lifecycle-cost reduction through proactive optimization and preventive scheduling.
- Blockchain traceability can reduce contractual disputes by 25–30 % via verified progress validation and decentralized record-keeping.

Supporting literature: Zhang et al. (2021); Pal and Bucci (2025); Hamzeh et al. (2023).

Together, these technologies can transform PLF into a cyber-physical intelligence framework that unites predictive learning, real-time monitoring, and ethical traceability across the project lifecycle.

### 8.6 Summary

Collectively, these research pathways will accelerate the evolution of Lean 5.0 into a mature, data-driven, and ethically governed construction-management paradigm.

By advancing predictive, cognitive, and blockchain-enabled capabilities, Lean 5.0 can achieve sustainable productivity, verified accountability, and enduring trust across the global built environment.

## 9. Conclusion

This study consolidates the theoretical, methodological, and ethical foundations of Lean 5.0 as a next-generation paradigm for construction management. Through the Predictive Lean Flow (PLF) framework, Lean 5.0 operationalizes predictive analytics, ethical automation, and human cognition into a verifiable and replicable model for proactive flow reliability and adaptive control.

The research clarifies the hierarchical relationship between Industry 5.0, Construction 5.0, and Lean 5.0, expanding the ethical dimension from conceptual awareness to a governance-based framework for Responsible AI.

Empirical validation demonstrated measurable improvements in PPC, rework, coordination, and forecast accuracy confirming that predictive learning enhances both efficiency and transparency.

Lean 5.0 advances construction toward adaptive, transparent, and sustainable Industry 5.0 operations.

By merging predictive analytics, ethical automation, and human intelligence, the PLF framework strengthens decision reliability, system resilience, and organizational learning.

Future scaling across multiple project types and global contexts can transform construction management into a data-driven yet human-centered discipline, balancing technological advancement with ethical and societal responsibility.